\documentclass[onecolumn,
 reprint,
 amsmath,amssymb,
 aps,
groupedaddress,superscriptaddress]{revtex4-2}

\usepackage{todonotes}
\usepackage{setspace}
\usepackage{cleveref}
\Crefname{equation}{Eq.}{Eqs.}
\crefname{pluralequation}{Eqs.}{eqs.}
\Crefname{figure}{Fig.}{Figs.}
\crefname{pluralfigure}{Figs.}{Figs.}
\Crefname{tabular}{Tab.}{Tabs.}

	\newcommand{\regist}{\textsuperscript{\textregistered}}

	\newcommand{\comsol}{COMSOL\regist}

	\newcommand{\mathematica}{Mathematica\regist}

	\newcommand{\gtheta}{\text{g}_{\theta}}
\usepackage{siunitx}
\usepackage{longtable}
\newcommand{\gn}{\text{g}_{N}}
\newcommand{\gcp}{\text{g}_{cp}}
\newcommand{\gcc}{\text{g}_{12}}
\newcommand{\cgg}{\text{C}_{1}}
\newcommand{\cng}{\text{C}_{2}}

\begin{document}




\title{Synchronization of silicon thermal free-carrier  oscillators}

\author{Gustavo de O. Luiz}
\affiliation{Instituto de Física Gleb Wataghin, Universidade Estadual de Campinas, Rua Sergio Buarque de Holanda, 777, Campinas-SP, Brasil}
\affiliation{nanoFAB Centre, University of Alberta, Edmonton, Alberta, T6G 2V4, Canada}

\author{Caique C. Rodrigues}
\author{Thiago P. M. Alegre}
\affiliation{Instituto de Física Gleb Wataghin, Universidade Estadual de Campinas, Rua Sergio Buarque de Holanda, 777, Campinas-SP, Brasil}
\author{Gustavo S. Wiederhecker}
\email[]{gsw@unicamp.br}
\affiliation{Instituto de Física Gleb Wataghin, Universidade Estadual de Campinas, Rua Sergio Buarque de Holanda, 777, Campinas-SP, Brasil}




\begin{abstract}
Recent exploration of collective phenomena in oscillator arrays has highlighted its potential for accessing a range of physical phenomena, from fundamental quantum many-body dynamics~\cite{PhysRevLett.107.043603} to the solution of practical optimization problems using photonic Ising machines~\cite{Marandi2014, PhysRevA.88.063853}. Spontaneous oscillations often arise in these oscillator arrays as an imbalance between gain and loss. Due to coupling between array individuals, the spontaneous oscillation is constrained and lead to interesting collective behavior, such as synchronized oscillations in optomechanical oscillator arrays~\cite{PhysRevLett.115.163902}, ferromagnetic-like coupling in delay-coupled optical parametric oscillators~\cite{Okawachi:15} and binary phase states in coupled laser arrays~\cite{Utsunomiya:15}. A key aspect of arrays is not only the coupling between its individuals but also their compliance towards neighbor stimuli. One self-sustaining photonic oscillator that can be readily implemented in a scalable foundry-based technology is based on the interaction of free-carriers, temperature and optical field of a resonant silicon photonic microcavity~\cite{Barclay:05, PhysRevLett.112.123901, Navarro-Urrios2017, ArreguiColombano, Borghi:21}. Here we demonstrate that these silicon thermal-free-carrier oscillators are extremely compliant to external excitation and can be synchronized up to their 16$^\text{th}$ harmonic using a weak seed. Exploring this unprecedented compliance to external stimuli, we also demonstrate robust synchronization between two thermal free carrier oscillators.
\end{abstract}


\maketitle

\section{Introduction}

Coupled oscillator phenomena plays an important role in physical systems, for instance, coupled oscillator arrays may lay out the synchronous behavior of complex biological systems~\cite{pikovsky2001synchronization}, or even emulate the behavior of coupled spins in solids that quickly solves complex optimization problems at low power consumption~\cite{PhysRevApplied.4.024016}. The dramatic evolution of micro and nanofabrication techniques has enabled the demonstration of these coupled oscillators using distinct physical systems such as lasers, optical parametric oscillators~\cite{Marandi2014}, and optomechanical oscillators~\cite{PhysRevLett.109.233906,Rodrigues2021}. In these examples, the oscillation emerges as a spontaneous process resulting from the imbalance between gain and loss around a resonant optical mode. When building coupled arrays of these oscillators, a major challenge is to ensure that their dissimilar resonant frequencies match each other – a constraint that often requires control loops~\cite{doi:10.1126/science.aav7932} or precise individual tuning. This is a consequence of their rather stiff oscillation frequencies that hardly deviate from the bare frequencies set by geometrical boundary conditions~\cite{PhysRevLett.109.233906,PhysRevLett.111.213902}.

More recently, a simple yet powerful oscillator has been explored in silicon, based on the interaction between the optical field, free charge carriers (FC), and temperature ~\cite{Johnson:06,PhysRevA.85.053819,PhysRevLett.112.123901}. The resulting oscillator frequency is not only very dependent on the intra-cavity energy, but is also very nonlinear, easily generating tens of harmonics~\cite{PhysRevLett.112.123901}. Such compliance of the oscillation frequency to intra-cavity energy has been recently explored in the synchronization of this oscillator to an optomechanical oscillator~\cite{Navarro-Urrios2015}.

In this work, we experimentally demonstrate that silicon free-carrier thermal oscillators exhibit exceptional compliance to external excitations and can be injection-locked up to their 16$^\text{th}$ harmonic using a weak modulation of the continuous wave pump laser. Leveraging this remarkable compliance, we also demonstrated the synchronization of a pair of coupled oscillators through their evanescent optical field. Furthermore, the compatibility of these oscillators with foundry-based CMOS technology should facilitate their scalability to large oscillator arrays.

\section{Thermal free carrier oscillators} 

Although silicon is transparent for light with wavelengths longer than approximately 1.1~\textmu m~\cite{https://doi.org/10.1002/pip.4670030303},  free-carriers (FC) are generated through two-photon absorption (TPA)~\cite{doi:10.1063/1.1571665,doi:10.1063/1.1435801} at large optical  intensities, leading to changes in the real and imaginary parts of the refractive index~\cite{6051462}. In an optical cavity, an increase in FC density (N) causes a blue shift of the resonance frequency ($\delta \omega$) due to a reduction in the refractive index ($\delta n$),as well as a nonlinear optical loss. Additionally, the temperature increase ($\theta$) resulting from linear and nonlinear optical absorption causes an increase in the refractive index due to the thermo-optic effect~\cite{PhysRevB.36.4821}, thereby red-shifting the optical resonance.

\begin{figure}[htbp]
    \centering
    \includegraphics[width=\linewidth]{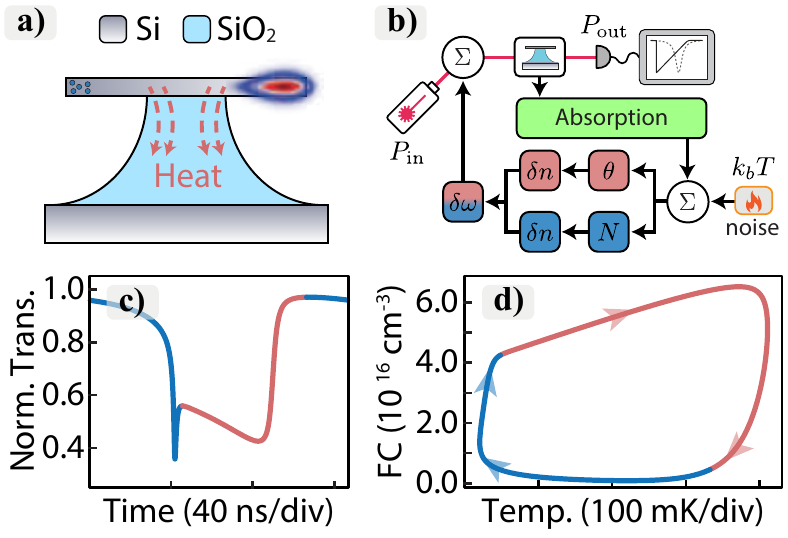}
    \caption{\textbf{a)} Schematic of the suspended silicon resonator, the colormap represent the optical field that generates heat and free charge carriers (represented by the tiny blue circles), higher field intensity in the red shading; \textbf{b)} Pictorial model of thermal-free carrier dynamics and closed loop diagram of the oscillating system. $P_{in}$: Input laser power, $P_{out}$: Output laser power; \textbf{c)} Simulated time trace of the transmitted signal; blue (red) parts indicate when the optical resonance is suffering blue (red) shifts in the cycle; \textbf{d)} Simulated dynamical FC density versus temperature. The closed cycle indicates self-sustained oscillations, the arrows indicate the time evolution.}
    \label{fig:1}
\end{figure}

As indicated in \cref{fig:1}(a), the optical field is concentrated on the disk's edge, where it is absorbed and generates FC and heat. The generated FC have typical life-times on the order of nanoseconds~\cite{Turner-Foster:10,Aldaya:16}, most likely due to surface facilitated recombination~\cite{doi:10.1063/1.1866635}. Heat, conversely, is dissipated mainly through the supporting silicon oxide pedestal, which has poor heat conduction characteristics, leading to microsecond scale thermal lifetimes. These two effects cause opposite refractive index changes, which results in resonance frequency shifts. This frequency shift alters the amount of energy stored inside the cavity, closes the feed-back loop represented in \cref{fig:1}(b) and amplifies small thermal fluctuations ($k_b T$, where $k_b$ is the Boltzmann constant and $T$ ambient temperature), which causes the self-pulsing of the stored optical energy encoded into the cavity transmitted light.

The model for such interaction in a single resonator is well known~\cite{PhysRevA.86.063808,PhysRevLett.112.123901,Borghi:21} and reproduces all qualitative features of the transmitted signal time-traces. An example of the simulated time-dependent transmission is shown in \cref{fig:1}(c) (see supplementary information for details). The two distinct time-scales, due to temperature and FC relaxations, are highlighted as distinct colors in the traces shown in \cref{fig:1}(c,d); there is a fast FC dominated region (blue), followed by a slow thermal regime (red). Such interplay can be more clearly seen by plotting the resulting FC density versus temperature variation, which defines a closed path in \cref{fig:1}(d), indicating that self-sustained oscillation regime was reached.

The closed trajectory in \cref{fig:1}(d) also allows us to understand the different aspects of this oscillation dynamics. For a pump laser ($P_{in}$) blue-detuned relative to the cavity optical resonance and above the oscillation power threshold, there is a rapid increase in the FC density that blue-shifts the resonance. This blue-shift is large enough to leave the laser red-detuned relative to the instantaneous optical resonance frequency, resulting in the narrow dip observed in \cref{fig:1}(c). At this point in the cycle the cavity temperature slowly increases, red-shifting the optical resonance frequency, until the laser is again blue-detuned, forming the second broader dip in the transmission trace, and the cycle is restarted.

\begin{figure}
    \centering
    \includegraphics[width=\linewidth]{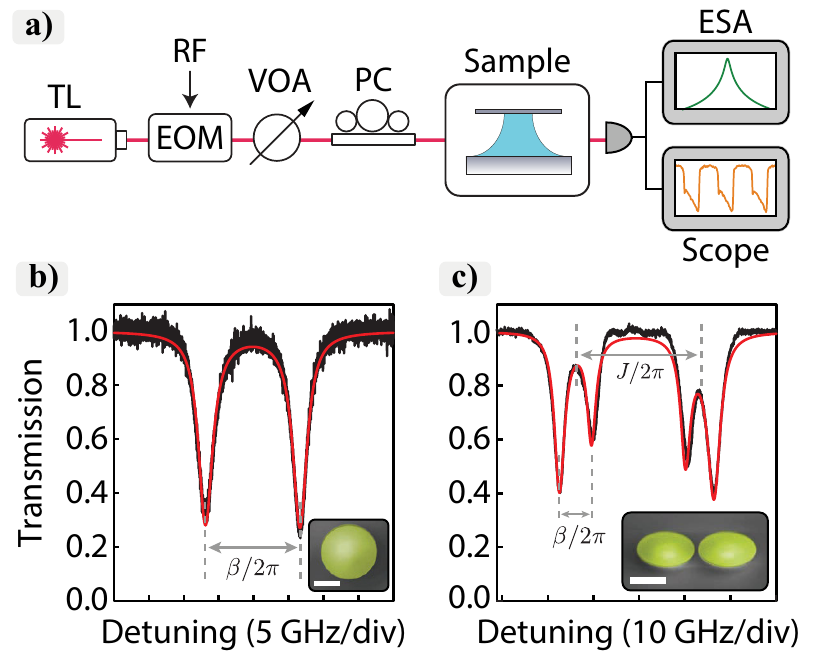}
    \caption{\textbf{a)} Simplified schematics of the experimental setup. TL: tunable laser; EOM: electro-optical amplitude modulator; VOA: variable optical attenuator; PC: polarization control; ESA: electrical spectrum analyzer; Scope: oscilloscope; \textbf{b)} Measured optical resonance of a single cavity. Center of resonances is at 1471.43~nm, with $\kappa/2\pi=2.5$~GHz linewidth and $\beta/2\pi=13.6~GHz$ splitting due to surface roughness; \textbf{c)} Measured optical resonance of the coupled resonator system. Center of the graph is at 1472.1~nm with linewidth $\kappa/2\pi=2.9$~GHz, intra-coupling strenght $\beta/2\pi=10~GHz$ and inter-coupling $J/2\pi\approx 28$~GHz. Inset shows false-colored SEM top view of each device, scale bars are 2~\textmu m long.}
    \label{fig:2}
\end{figure}

The optical properties of the two microdisk devices explored in this work are characterized in \cref{fig:2}(b) and \cref{fig:2}(c). For the single resonator device (2 \textmu m radius, 220 nm thick silicon), as shown in \cref{fig:2}(b), the optical resonance is centered at 1471.4~nm, with a loaded linewidth of $\kappa/2\pi=2.5$~GHz ($Q=81\times10^3$). Due to coupling of counter-propagating modes caused by surface roughness~\cite{Borselli:07}, two resonance peaks are observed split by $\beta/2\pi=13.6$~GHz. Similarly, the coupled-resonator device (2 \textmu m radii, 220 nm thick silicon, 300~nm coupling gap) has resonances centered at 1472.1~nm and shows 4 distinct peaks. The pair of peaks at each detuning sign are also due to coupling between counter-propagating modes ($\beta/2\pi=10$~GHz), while the splitting between the left and right-hand peak pairs is due to the evanescent coupling between the two cavities ($J/2\pi\approx 28$~GHz). The fitted linewidth for the coupled cavities is found to be $\kappa/2\pi=2.9$~GHz ($Q = 70\times 10^3$).

\section{Injection locking of a single oscillator}

We investigate the compliance of these oscillators by injection locking them to a weak  external signal (at a frequency $f_\text{inj}$) that was imposed on the amplitude of the pump laser. To achieve this modulation, an electro-optical amplitude modulator (EOM) is inserted in the light path before the resonators. A simplified schematic of the experimental setup is shown in \cref{fig:2}(a), consisting of a pump laser (TL, for tunable laser), a variable optical attenuator (VOA), a polarization controller (PC) and a photo-detector. The pump laser is fiber coupled, and a tapered fiber is used to evanescently couple the light into the cavity~\cite{knight1997}. The transmitted light is detected, and the slow (DC) component of the photocurrent monitors the average cavity transmission while the fast (AC) components are sent to an oscilloscope and to an electrical spectrum analyzer (ESA). Additionally, we used a calibrated acetylene gas cell and a Mach-Zehnder interferometer for calibrating the laser frequency, which are not shown in the figure.

\begin{figure*}[ht]åç
    \centering
    \includegraphics[width=\linewidth]{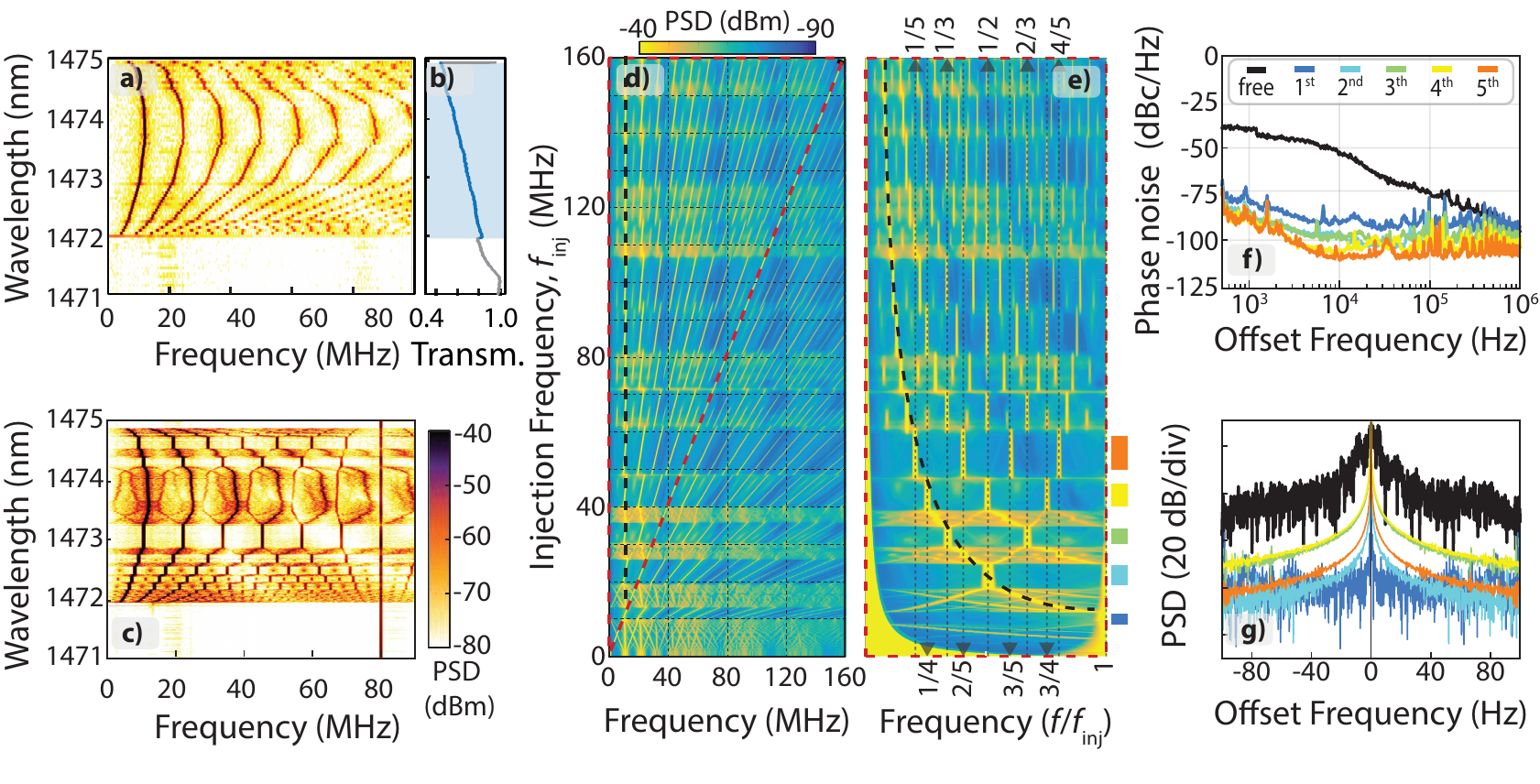}
    \caption{ \textbf{Injection locking of a single resonator.} \textbf{a)} Single-cavity transmission radio-frequency spectrogram as the continuous-wave laser wavelength is swept around an optical resonance, as measured by an ESA; \textbf{b)} DC optical transmission measured simultaneously with the spectrogram in part (a). Blue-shaded region indicates thermal free-carrier self-sustaining oscillations; \textbf{c)} Single-cavity transmission radio-frequency spectrogram, similar to (a), but with an amplitude-modulated pump serving as an injection signal at $f_\text{inj}=80$~MHz; \textbf{d)} Raw spectogram obtained at fixed pump wavelength (1473 nm) while the injection frequency ($f_\text{inj}$) is swept. The red-dashed triangle region highlights the high-harmonic synchronization region, where the injection frequency is larger than the oscillator frequency, the vertical black-dashed line indicates the fundamental frequency of the oscillator.  \textbf{e)} The same spectrogram as in (d), but with the frequency axis rescaled by the injection signal frequency. In this rescaled diagram, the red-dashed triangle region of (d) is mapped into the rectangular region, the vertical black-dashed line indicates the fundamental oscillator tone; thin vertical-dashed lines are guides-to-the-eye for the harmonic relations ($p/q$); \textbf{f)} Phase noise measurement at the fundamental oscillator tone while it is free-running and injection-locked at the first 5 harmonics, the corresponding regions where phase-noise is measured is highlighted by the colored rectangles in the lower right of (d); \textbf{g)} First harmonic spectrum when system is free-running and locked at the first 5 harmonics. The colors match part (f).}
    \label{fig:3}
\end{figure*}

A simple laser wavelength sweep across the optical resonance of the single resonator device, without the amplitude modulation, already reveals the oscillation dynamics described in \cref{fig:1}. With an optical input power of only 310~\textmu W, as the laser reaches a threshold detuning (highlighted in blue in the transmission plot of \cref{fig:3}(b)), the radio-frequency spectrogram captured by the ESA reveals multiple harmonics of the transmitted signal (x-axis is the ESA calibrated frequency). Moreover, a large optical spring effect is observed as the frequency of the oscillator varies with the laser wavelength (y-axis in \cref{fig:1}(a,c)), which is a hint that the pulses period and duty-cycle (ratio between duration and period) strongly depend on the optical energy stored in the resonator~\cite{PhysRevLett.112.123901}.

Injection locking can be readily observed by driving the EOM with a weak $f_\text{inj}=80$~MHz sinusoidal electrical drive (drive power of -10~dBm, modulation depth of 8\%). In the presence of the modulation signal, the laser sweep through the resonance results in a dramatically distinct RF spectrogram, as shown in \cref{fig:3}(c). Although a wide spring effect is still observed, as in \cref{fig:3}(a), the resonator frequency evolves in discrete steps, each corresponding to non-trivial fractional sub-harmonics of the RF tone, forming a sequence of the so-called devil staircases~\cite{PhysRevE.93.023007}. The first step of the devil staircase is observed while the oscillator first harmonic is at 5~MHz, which implies that it is being entrained by a modulation at its 16$^\text{th}$ harmonic.

To explore the injection locking behavior in greater detail, we presented the RF spectrogram in \cref{fig:3}(d), with the x-axis representing the ESA calibrated frequency and the y-axis representing the linearly swept modulation injection frequency from zero to 160~MHz, while the laser is fixed at a certain wavelength. The spectrogram in \cref{fig:3}(d) reveals multiple synchronization regions, identifiable by the sloping harmonic lines to the right, rather than vertical lines. Such sloped harmonics indicate the fractional ratios where the oscillator is entrained. To simplify the  identification of the fractional ratios, we normalized the horizontal frequency axis by dividing it by the injection frequency (vertical axis in \cref{fig:3}(d)). The resulting frequency-normalized map is presented in \cref{fig:3}(e), with the same y-axis as in \cref{fig:3}(d), which clearly shows the precise fractional synchronization orders ($p/q$, where $p$ is the observed harmonic and $q$ is the externally driven harmonic). In this representation, the extent of the locking ranges is also noticeable as vertical lines, corresponding to several MHz even at very high harmonics. Such robust fractional synchronization orders are considered rare~\cite{pikovsky2001synchronization}, but not in such thermal free-carrier oscillators. Such broad high-order synchronization windows have been observed in optomechanical systems, but only up to the 4$^\text{th}$ harmonic~\cite{Rodrigues2021}. The greater compliance observed in these free-carrier thermal oscillators can be attributed to their resemblance to relaxation oscillators. The reason for this is the rapid relaxation time-scales exhibited by thermal-free carriers, which produce an oscillator output similar to a pulse. This results in a broader frequency response when compared to optomechanics, where the response tends to be more sinusoidal due to the resonant filtering of the mechanical modes. These nonlinear features of the thermal-free carrier limit cycle have been extensively discussed by Abrams et al.~\cite{PhysRevLett.112.123901}.

A useful characteristic of high-harmonic synchronization is the possibility to effectively suppress the oscillator phase-noise (PN) characteristics~\cite{Leijssen2017,6043392}. The suppression of phase-noise, however, relies on the compliance of the oscillator to the injected signal. The wide locking ranges observed in figures \ref{fig:3}(d,e) suggests that our oscillators are strong candidates to observe such phase-noise suppression. Indeed, as we measure the phase-noise performance when the oscillator is entrained at different harmonics, we observe a consistent suppression of phase-noise at the oscillator's fundamental tone. \Cref{fig:3}(f) shows the measured PN spectra of the fundamental tone when the system is free-running (no modulation) and when it is injection-locked up to the fifth harmonic ($p/q=1/(1,2,...,5)$), demonstrating a clear PN suppression. The spectral characteristics of the fundamental tone, which correspond to the PN measurements in \cref{fig:3}(f), are depicted in \cref{fig:3}(g). The wide locking range observed for the $p/q=1/5$ excitation in \cref{fig:3}(e) suggests that this excitation should have the best phase-noise suppression, in line with usual phase-noise modeling~\cite{plessas} and previous studies~\cite{plessas,Rodrigues2021}. The overall suppression of PN for higher harmonic injection is also expected, as shown in \cref{fig:3}(f).

Such wide locking ranges are advantageous when exploring synchronization between two or more oscillators, even if they differ slightly due to fabrication fluctuations. Moreover, the fact that locking can happen at different harmonics also permits that two oscillators synchronize at different frequencies, an unusual characteristic that cannot be found in sinusoidal oscillators~\cite{pikovsky2001synchronization}. Furthermore, high-harmonic synchronization could be explored for all-optical radio-frequency division, a fundamental towards frequency synthesis~\cite{Rodrigues2021}.

\section{Synchronization of coupled oscillators}

\begin{figure*}[ht]
    \centering
    \includegraphics[width=\linewidth]{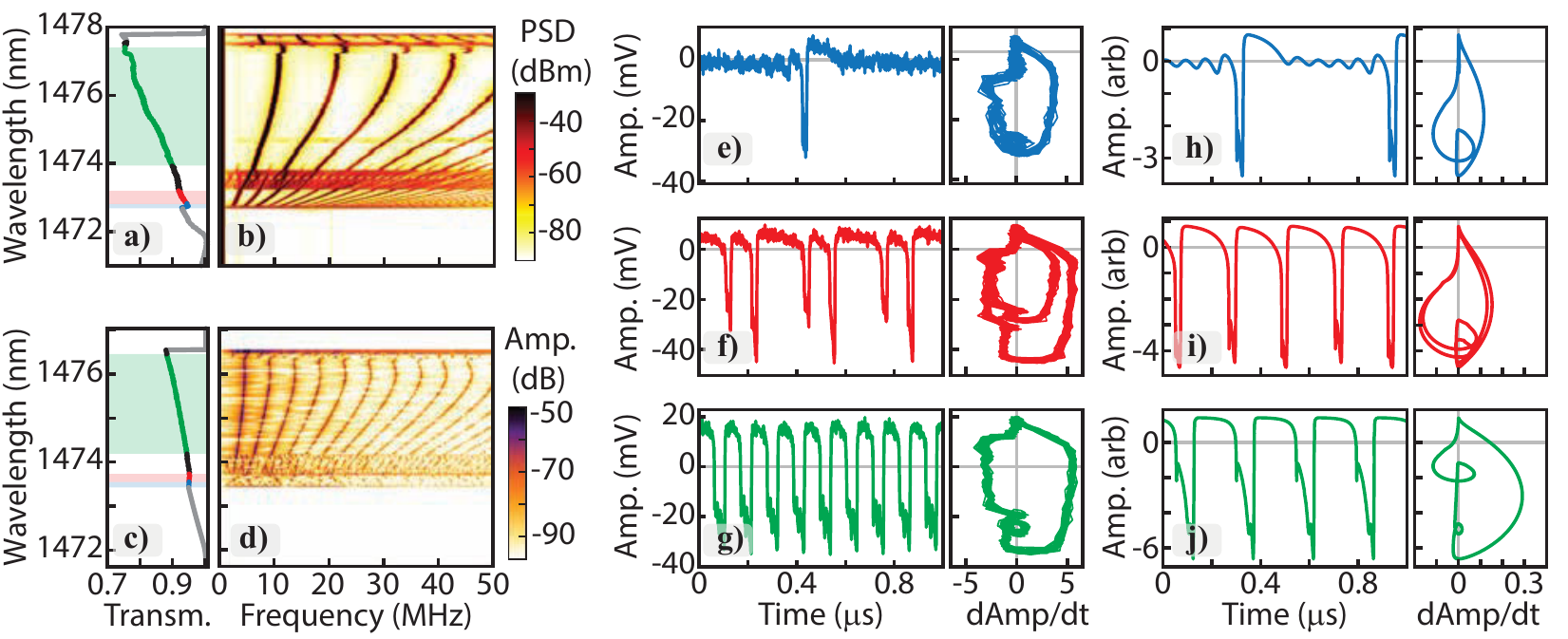}
    \caption{ \textbf{Synchronization of two oscillators}. \textbf{a)} Experimental DC optical transmission of self-oscillating coupled resonators; \textbf{b)} Experimental wavelength varying spectrogram of coupled resonators; \textbf{c)} Numerical DC optical transmission of self-oscillating coupled resonators; \textbf{d)} Numerical wavelength varying spectrogram of coupled resonators. In (a, c), colored regions mark different regimes of operation. \textbf{e-g)} Experimental time traces at different modes of operations; \textbf{h-j)} Numerical time traces at different modes of operation. In (e-j), colors match the regions in (a, c).}
    \label{fig:4}
\end{figure*}

The injection locking results reveal a remarkable property of thermal free carrier oscillators: their extreme compliance to external stimuli. In this section, this property is explored to demonstrate synchronization between two oscillators coupled by their evanescent optical field, (see inset in \cref{fig:2}(c)). Without employing any external modulation, the pump laser is swept, in steps of 50 pm, from shorter to longer wavelength around a set of selected optical resonances (shown in \cref{fig:2}(c)). The pump laser power was increased to approximately 2~mW to ensure that the oscillation threshold is reached, a consequence of the larger optical mode volume and weaker coupling of each optical mode to the bus waveguide (tapered optical fiber).

Instead of observing a steady evolution of the oscillator frequency, as in \cref{fig:3}(a), the coupled cavity dynamics exhibits a more complex behavior. Such behavior, shown in \cref{fig:4}(a,b) is very well reproduced by our numerical model (\cref{fig:4}(c,d), see supplementary information). The first striking difference we observe in the double cavity system spectrogram is the appearance of qualitatively distinct regions (highlighted in blue,red and green in \cref{fig:4}(a,c)). The temporal behavior of the transmitted signal (AC component) measured by the oscilloscope is shown in \cref{fig:4}(e-g); along with their simulation counterparts in \cref{fig:4}(h-j). When the system is just starting to oscillate (blue region in \cref{fig:4}(a,c)) a single low frequency pulse appears, similar to the single cavity oscillation condition. In this case there is a single path in both measured and simulated phase diagrams (amplitude versus amplitude's time derivative), as shown in \cref{fig:4}(e,h)).

As the pump reaches the red region in \cref{fig:4}(a,c), a second distinct peak appears in the time trace (\cref{fig:4}(f,i)) and the spectrogram in \cref{fig:4}(b) shows very clear harmonics. Also, the phase diagram presents two distinct paths. This is the first feature we observe in the time trace that is exclusive of the double cavity system. Further red-shifting the pump wavelength, in between the red and green regions in fig.\cref{fig:4}(a), the two oscillators lose synchronization and produce erratic pulses, which manifests as an elevated noise level in the spectrogram. Finally, once the green region is reached, the two oscillators reestablish synchronization for a wide range of laser detuning ($\approx3.8$~nm).

Although the experimentally observed features suggest that the coupled oscillator dynamics and synchronization are present, it is difficult to grasp the dynamics of temperature, free-carrier density, and light from the transmission traces alone. To better understand the dynamics in each of the indicated regions, we now turn our attention to the numerical simulations of the coupled system.

\subsection{The intricate dynamics of synchronized oscillators}

The good qualitative agreement with the theoretical model highlighted in \cref{fig:4} suggests that individual dynamics may be unveiled by exploring the numerical model. In particular, we extract the instantaneous optical frequency shift ($\Delta\omega_{1,2}(t)$) of each cavity based on their coupling to the time-dependent free-carrier density ($\text{N}_{1,2}(t)$) and temperatures variation ($\theta_{1,2}(t)$),
\begin{align}
    \Delta\omega_1(t) &=g_\text{th}\theta_1(t) +g_\text{N} N_1(t) \label[pluralequation]{eq:fshift1}  \\
    \Delta\omega_2(t) &=g_\text{th}\theta_2(t) +g_\text{N} N_2(t),
    \label[pluralequation]{eq:fshift2}
\end{align}
\noindent where $g_\text{th}$ and $g_\text{N}$ are the thermal and carrier coupling parameters; $\theta_{1,2}(t)=T_{1,2}(t)-T_0$ (see supplementary for definition of all terms). As detailed in the supplementary material, cavity 1 is the one coupled to the waveguide, while cavity 2 only receives light through the coupling to its companion.

\begin{figure}[ht]
    \centering
    \includegraphics[width=\linewidth]{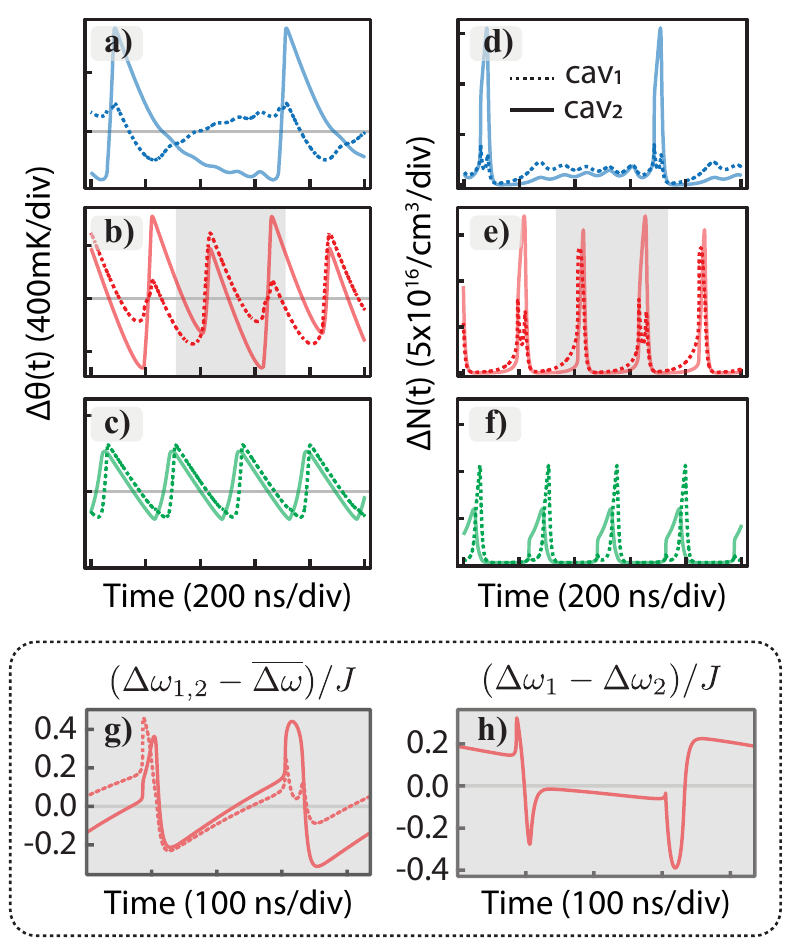}
    \caption{ \textbf{Temperature variations and FC density of a self-pulsing double optical cavity system.} \textbf{a-c)} AC temperature variation of each resonator; \textbf{d-f)} AC free-carrier density variation of each resonator. The AC part of each contribution is obtained by subtracting the time-average of each temperature/carrier dynamical variable; \textbf{g)} Individual cavity resonance shifts around equilibrium ($\overline{\Delta\omega}$); \textbf{h)} Difference between the two cavities' resonances. In (g, h) the frequencies have all been normalized by the inter-cavity coupling rate. In all figures, the colors correspond to the traces show in \cref{fig:4}(h-j)}
    \label{fig:5}
\end{figure}

We investigate these parameters to show that the temporal characteristics revealed in \cref{fig:4}(h-j) are caused by carrier and temperature dynamics that are not always synchronized in the trivial ($p/q=1/1$) situation. Moreover, the frequency oscillations ($\Delta\omega_{1,2}(t)$) are comparable to the evanescent coupling rate $J$ and affect the coupling between the optical modes. The individual AC components of temperature ($\Delta \theta_{1,2}(t)=\Delta \theta_{1,2}(t)-\overline{\theta}_{1,2}$) and FC contributions ($\Delta N_{1,2}(t)=N_{1,2}(t)-\overline{N}_{1,2}$, where the bar indicates average over time) to the resonator frequencies are shown in \cref{fig:5}(a-c) and \cref{fig:5}(d-f), with colors corresponding to the transmission traces highlighted in \cref{fig:4}(h-j). \Cref{fig:5}{(a,d)} reveals that the single cycle behavior observed in \cref{fig:4}(e,h) occurs when the dynamics is concentrated only in cavity 2, as both $\Delta \theta_{1}$ and $\Delta N_{1}$ are significantly smaller in resonator 1. As a result, the time trace is similar to the single oscillator case.

In the red region shown in \cref{fig:5}(b,e), both oscillators display similar magnitudes. However, we observed that cavity 2 consistently exhibits peaks in the carrier density $\Delta N_{2}$, while cavity 1 appears to skip every other pulse. This indicates synchronization between the two cavities, but not at the trivial condition of $p/q=1/1$. Rather, cavity 2 oscillates at twice the frequency of cavity 1 ($p/q=1/2$). We also observed similar behavior for $\Delta \theta_{2}$, albeit with a smaller contrast between the two cavities. The RF spectrograms in this region (\cref{fig:4}(b)) reflect this temporal behavior by showing a denser pattern of harmonics. This characteristic is further highlighted in \cref{fig:5}(g), where we examine the behavior of the time-dependent frequency shifts (\cref{eq:fshift1,eq:fshift2}) in the gray-shaded region of \cref{fig:5}(b,e). During the first pulse, both cavities experience similar resonance frequency shifts, while during the second pulse, only cavity 2 undergoes a significant shift.

Another important aspect is that the magnitude of the frequency shift is comparable with the evanescent coupling rate $J$, showing that the very coupling between the cavities is affected by their dynamical behavior. Indeed, we show in \cref{fig:5}(h) the difference between the cavities' resonances, showing that their relative frequency changes as much as $\Delta\omega_1-\Delta\omega_2\approx 0.4J$. Such large changes in the bare resonator frequencies significantly affect the energy stored in each cavity and, thus, the oscillation dynamics.

Finally, in the green region (\cref{fig:5}(c, f)), where the synchronization is very well defined and the pulsing behavior is regular in \cref{fig:4}(g, j), both temperature (\cref{fig:5}(c)) and free-carrier density (\cref{fig:5}(f)) oscillate with the same period. Nevertheless, even though the temperature behavior of both resonators only shows a small offset, the free-carrier density presents quite a different behavior. The slightly delayed peak in cavity 1 causes a quick blue-shift in its optical resonance frequency, while cavity 2 is already finalizing its cycle and red-shifting. This is what causes the extra knot we observe in the phase diagrams in \cref{fig:4}(j).

\section{Conclusion}

In summary, we demonstrated that carrier-thermal oscillations are extremely compliant to external stimuli, and as such can be synchronized even by weak signals at a wide range of its harmonics. The low phase-noise performance and the multi-octave spanning synchronization range could be explored for all-optical RF frequency division. Furthermore, we explored the extreme compliance of such oscillator and demonstrated the intricate synchronization dynamics of a pair of such resonators. The numerical model of coupled resonators allowed a detailed understanding of the intricate synchronization dynamics. These synchronized oscillators, albeit nominally identical exhibit synchronization at distinct harmonics, opening a path for scalable array synchronization of dissimilar silicon oscillators.


\paragraph*{Funding}
This work was supported by São Paulo Research Foundation (FAPESP) through grants
19/14377-5, 
18/15577-5, 
18/15580-6, 
18/25339-4, 
Coordenação de Aperfeiçoamento de Pessoal de Nível Superior - Brasil (CAPES), and CNPq.

\paragraph*{Acknowledgments}
The authors thank A. Von Zuben for technical support. We also acknowledge CCSNano-UNICAMP for providing part of the microfabrication infrastructure.
\paragraph*{Disclosures}
The authors declare no conflicts of interest.

\paragraph*{Data Availability}
All simulation files and data are available upon request to the authors.
\bibliography{bibliography}

\makeatletter
\renewcommand{\thesection}{S}
\makeatother

\counterwithin{figure}{section}
\counterwithin{table}{section}
\onecolumngrid
\appendix
\section{Supplementary Information\label{supp}}

\subsection{Theoretical model}
For a single cavity, we model this phenomenon with three coupled equations, one for the optical field (eq.~\ref{eq:5.1}), one for the density of charge carriers (eq.~\ref{eq:5.2}) and one for the temperature variation (eq.~\ref{eq:5.3})~\cite{PhysRevA.86.063808,PhysRevLett.112.123901}:
\begin{align}
\frac{d a(t)}{d t} &= i \left[ \Delta + \gtheta \theta(t) + \gn N(t)  \right] a(t) - \frac{\kappa+\alpha_{TPA} \left|~a(t)~\right|^2 + \alpha_N \left| N(t) \right|}{2} a(t) + \sqrt{\kappa_e P_{in}}\label{eq:5.1}\\
\frac{d N(t)}{d t} &= -\gamma_{FC} N(t) + \beta_{FC} \left|~a(t)~\right|^4\label{eq:5.2}\\
\frac{d \theta(t)}{d t} &= -\gamma_{th} \theta(t) + \beta_{th} \left( \kappa_{lin} + \sigma_{Si} v_g N(t) + \alpha_{TPA} \left|~a(t)~\right|^2 \right) \left|~a(t)~\right|^2\label{eq:5.3}
\end{align}

The definition of the parameters of these equations are summarized on table~\ref{tab:5.1}.

\begin{longtable}[ht!]{|c|m{4.50cm}||c|m{3.20cm}|}
	\caption{\label{tab:5.1}\textbf{Parameters of the self-pulsing equations.}}
	\renewcommand{\arraystretch}{1.4}
	\setlength\tabcolsep{1pt}
		\tabularnewline
		\hline
		\textbf{Parameter} & \centering \textbf{Description} & \textbf{Parameter} & \centering \textbf{Description} \tabularnewline \hline
		$\left|~a~\right|^2$ & Intra-cavity stored energy & $\gamma_{FC}$ & Free-carriers decay rate\\ \hline
		$N$ & Density of carriers & $\gamma_{th}$ & Thermal decay rate\\ \hline
		$\theta = T-T_0$ & Temperature variation with respect to equilibrium & $\gtheta$ & Coupling coefficient of optical resonance to temperature variation\\ \hline
		$\Delta = \omega_l-\omega_0$ & Pump laser frequency detuning with respect to the optical resonance of the cavity & $\gn$ & Coupling coefficient of optical resonance to the density of carriers\\ \hline
		$\kappa$ & Total linear decay rate of the optical cavity (includes linear absorption, scattering and coupling to the waveguide) & $\alpha_{TPA}$ & Two-photon absorption loss parameter\\ \hline
		$\kappa_e$ & Coupling rate between optical cavity and waveguide & $\alpha_N$ & Free-carrier optical loss parameter\\ \hline
		$\kappa_{lin}$ & Optical decay rate due to linear absorption & $\beta_{FC}$ & Free-carrier generation parameter\\ \hline
		$v_{g}$ & Optical mode group velocity & $\beta_{th}$ & Thermal source parameter\\ \hline
		$\sigma_{Si}$ & Silicon free-carrier absorption cross-section & $ P_{in} $ & Pump laser power\\ \hline
\end{longtable}

The coupling coefficients of the optical resonance to temperature and carrier density are given by
\begin{equation}
\gtheta = \frac{\omega_0}{n_g} \frac{d n}{d T}\label{eq:5.4}
\end{equation}
and
\begin{equation}
\gn = \frac{\omega_0}{n_g}\frac{d n}{d N}, \label{eq:5.5}
\end{equation}
\noindent
where $n$ is the refractive index, $ n_{g} $ is the group index of refraction, $\omega_0$ is the optical resonance angular frequency in the absence of non-linear effects (cold-cavity) and $T$ is the temperature. Note that the opposite effect of $ \theta $ and $ N $ is implicit in this equations through $ dn/dT $, which is positive, and $ dn/dN $, which is negative.

The TPA and FC absorption parameters are defined as
\begin{equation}
\alpha_{TPA} = \frac{\Gamma_{TPA}\beta_{Si} c^2}{V_{TPA} n_g^2}\label{eq:5.6}
\end{equation}
and
\begin{equation}
\alpha_N = \frac{\sigma_{Si} c}{n_g},\label{eq:5.7}
\end{equation}
\noindent
where $\beta_{Si}$ is the TPA constant for silicon and $c$ is the speed of light.

The FC and thermal source parameters are defined as
\begin{equation}
\beta_{FC} = \frac{\Gamma_{FC} \beta_{Si} c^2}{2 \hbar \omega_0 n_g^2 V_{FC}^2}\label{eq:5.8}
\end{equation}
and
\begin{equation}
\beta_{th} =\frac{\Gamma_{disk}}{\rho c_p V_{disk}},\label{eq:5.9}
\end{equation}
\noindent
where $\hbar$ is the reduced Planck constant and $c_p$ is the material heat capacity. The parameters $\Gamma_{TPA}$ and $\Gamma_{FC}$ are overlap factors for TPA and FC absorption, respectively, while $V_{TPA}$, $V_{FC}$ and $V_{disk}$ are effective volumes available for TPA and FC absorption and the total disk volume, respectively. These parameters values are determined from finite element method (FEM) numerical simulations performed using~\comsol~and are defined as~\cite{Barclay:05}
\begin{align}
\Gamma_{TPA} &= \frac{\int_{Si}n^4\left(\vec{r}\right)E^4\left(\vec{r}\right)d\vec{r}}{\int n^4\left(\vec{r}\right)E^4\left(\vec{r}\right)d\vec{r}}\text{,} \label{eq:5.10}\\
\Gamma_{FC} &= \frac{\int_{Si}n^6\left(\vec{r}\right)E^6\left(\vec{r}\right)d\vec{r}}{\int n^6\left(\vec{r}\right)E^6\left(\vec{r}\right)d\vec{r}}\text{,}\label{eq:5.11}\\
V_{TPA} &= \frac{\left(\int n^2\left(\vec{r}\right)E^2\left(\vec{r}\right)d\vec{r}\right)^2}{\int n^4\left(\vec{r}\right)E^4\left(\vec{r}\right)d\vec{r}} \label{eq:5.12}
\end{align}
and
\begin{equation}
V_{FC}^2 = \frac{\left(\int n^2\left(\vec{r}\right)E^2\left(\vec{r}\right)d\vec{r}\right)^3}{\int n^6\left(\vec{r}\right)E^6\left(\vec{r}\right)d\vec{r}}, \label{eq:5.13}
\end{equation}
\noindent
where $ E(\vec{r}) $ is the spacial distributions of the electric field amplitude.

There are 3 distinct time scales present in equations~\ref{eq:5.1} to~\ref{eq:5.3}. The fastest time-scale is the optical life-time ($\tau_{opt} = 1/\kappa$), which is in the order of tens of picoseconds for cavities with optical quality factors on the order of 80k. Then comes the FC life-time ($\tau_{FC} = 1/\gamma_{FC}$), in the order of a few nanoseconds. And the slowest time-scale is the thermal life-time ($\tau_{th} = 1/\gamma_{th}$), in the microsecond scale. One can then say that the optical field inside the cavity responds instantaneously to any FC or thermal variation. This means that the stored optical energy ($ \left|~a~\right|^2 $) is always in steady-state at any time, following any variations of $N$ and $\theta$ adiabatically. We can then write $da/dt = 0$ and solve the first equation for $a(t) = a(N(t),\theta(t))$.

\begin{figure}[ht!]
	\center
	\includegraphics[scale=1.0]{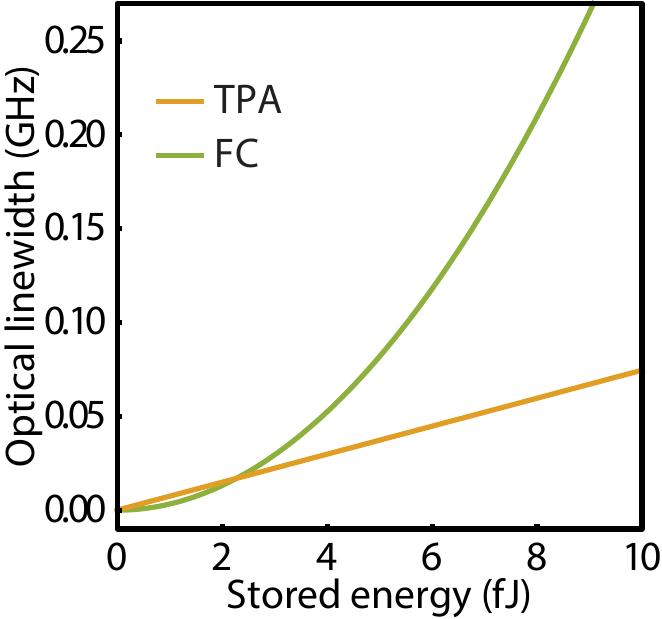}
	\caption[Optica losses due to FC absorption and TPA as a function of the intra-cavity energy]{\label{fig:losses_energy}\textbf{Optical losses due to FC absorption and TPA as a function of the intra-cavity energy.}}
\end{figure}

To do so it is necessary first to take care of a non-linear absorption term that depends explicitly on $a(t)$ in equation~\ref{eq:5.1}, namely the optical loss due to TPA. In fact, by solving the steady-state problem for the carriers density ($ dN(t)/dt=0 $) it is possible to evaluate the contribution of each loss term in equation~\ref{eq:5.1} at any given static amount of energy inside the cavity (fig.~\ref{fig:losses_energy}). For a typical input power of 1~mW to 2~mW, the stored energy inside of our cavities is on the order of 10~fJ. For this amount of energy both non-linear absorption coefficients are still smaller than the linear one ($ \kappa $), but the terms due to TPA is much smaller than the other two, unless for very little amounts of stored energy when it is comparable to the FC absorption term.

Nevertheless, because the TPA term is always much smaller than the total optical decay rate, this term can be ignored and the steady-state solution of equation \ref{eq:5.1} can be simplified, resulting in a single and simple solution for the intra-cavity field amplitude, given by
\begin{equation}
\label{eq:5.14}
a(t) = \frac{2 \sqrt{\kappa_{e} P_{in}}}{2 i \left(\Delta + + \gtheta \theta(t) + \gn N(t)  \right) - \kappa + \alpha_N \left| N(t) \right|}.
\end{equation}

Inserting equation~\ref{eq:5.14} into equations~\ref{eq:5.2} and~\ref{eq:5.3} we obtain two equations of real arguments for the temperature variation and FC density, given by
\begin{align}
\frac{d N(t)}{d t} &= -\gamma_{FC} N(t) + \beta_{FC} U(t)^2\label{eq:5.15}\\
\frac{d \theta(t)}{d t} &= -\gamma_{th} \theta(t) + \beta_{th} \left( \kappa_{lin} + \sigma_{Si} v_g N(t) + \alpha_{TPA} U(t) \right) U(t),\label{eq:5.16}
\end{align}
\noindent
where $ U(t)=\left|~a(t)~\right|^2 $, for simplicity. This approximation will be of special value when solving the problem of 2 cavities, as it will lead to a smaller memory consumption and shorter time to solve the problems. It is important to note that, although the TPA term was removed from the optical equation (eq.~\ref{eq:5.1}), this term is maintained in equation~\ref{eq:5.3}, minimizing the amount of approximations to solve the problem.

The material parameters used to properly solve the problem are obtained from the literature~\cite{Barclay:05,doi:10.1063/1.1713251,PhysRev.134.A1058,Asheghi1998,Borselli:07,Aldaya:16,Johnson:06}. The thermal life-time was estimated by considering the conduction of heat from the top silicon layer through a truncated cone oxide pedestal, with top diameter given by the undercut of the disk obtained from optical images and bottom diameter was made equal to that of the disk. This produced a typical time-scale of 7~\textmu s for the single cavity and 3.5~\textmu s for the double cavity, with the difference given by the different undercut size in each sample. The FC life-time for both systems was assumed to be 4 times the value presented in reference~\cite{Aldaya:16}, which results in a timescale of 5.8~ns. We chose reference~\cite{Aldaya:16} because their measurements were performed in waveguides fabricated by the same foundry as the samples used in this work. The scale factor of the FC lifetime was chosen so that the time dependent solutions had similar frequencies to the experimental data. The linear absorption rate ($ \kappa_{lin} $) was set to half of the intrinsic linear decay rate ($ \kappa_{i} $), based on the results presented in reference~\cite{Borselli:07}.

The material and other parameters used in the model are listed in  \cref{tab:params}. Here $ t_\text{Si} $ is the silicon layer thickness (220~nm) and $ r_{disk} $ is the disk radius (2~\textmu m). The two values given for $ \tau_{th} $ are for the double and single cavity systems, respectively.

\begin{table}
	\center
	\caption[]{\textbf{Table with parameter values used in the numerical simulations.}}
	\begin{tabular}{|c|c|}
		\hline
		\textbf{Parameter} & \textbf{Value}\\\hline
		$ \rho $ & 2330~kg/m$ ^{3} $\\
		$ c_{p} $ & 712~J/(kg K)\\
		$ n_{Si} $ & 3.485\\
		$ n_g $ & 0.99 $ n_{Si} $\\
		$\lambda_0$ & 1471.63~nm\\
		$\kappa$ & $ 2 \pi \times 2.55$~GHz\\
		$\kappa_e$ &  $ 2 \pi \times 1.17$~GHz\\
		$\kappa_i$ & $ \kappa - \kappa_{e} $\\
		$\kappa_{lin}$ & $ \kappa_{i}/2 $\\
		$ \tau_{FC} $ &  $ 5.8 $~ns\\
		$\gamma_{FC}$ & $ 1/\tau_{FC} $\\
		$ \tau_{th}$ & (7.0, 3.5)~\textmu s\\
		$\gamma_{th}$ & $1/\tau_{th}$\\
		$ T_0 $ & 300~K\\
		$dn/d\theta$ & $ 1.86\times 10^{-4} $~K$ ^{-1} $\\
		$dn/dN$ & $ -1.73 \times 10^{-27} $~m$ ^{3} $\\
		$\sigma_{Si}$ & $ 10^{-21} \text{m}^{2}$\\
		$ \beta_{Si} $ & $ 8.4 10^{-12} $~m/W\\
		$ \Gamma_{TPA} $ & 1\\
		$ V_{TPA} $ & $ 1.24\times10^{-18} \text{m}^{3}$\\
		$ \Gamma_{FC} $ & 1\\
		$ V_{FC} $ &  $ 1.15\times10^{-18} \text{m}^{3}$\\
		$ \Gamma_{disk} $ & 0.97\\
		$ V_{disk} $ & $ \pi r_{disk}^{2} t_{Si}  = 2.76\times 10^{-18} \text{m}^{3} $\\\hline
	\end{tabular}
	\label{tab:params}
\end{table}
\subsection{Numerical solution details}
\begin{figure}[ht!]
	\center
	\includegraphics[scale=1.0]{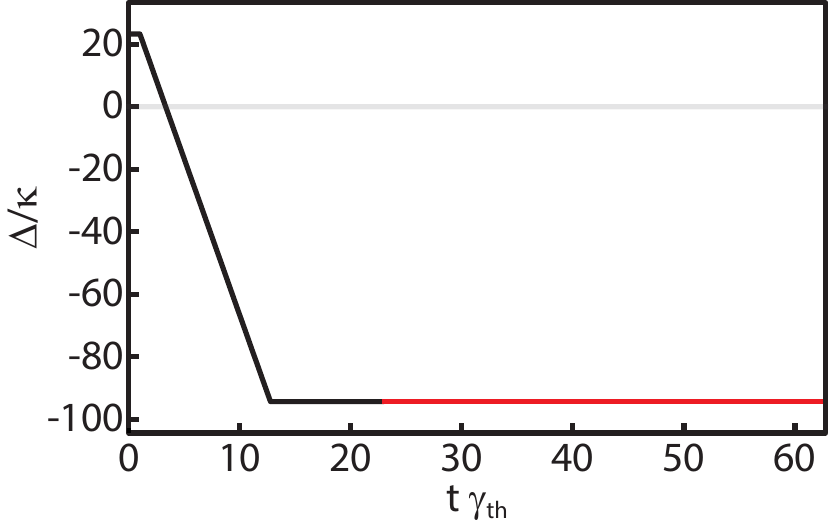}
	\caption[Time dependent detuning function.]{\label{fig:DetTime}\textbf{Time dependent detuning function.} The black region simulates the laser frequency sweep, taking the optical resonance non-linear shift into account. The red region is where time dependent data is stored and the Fourier transform evaluated. The horizontal axis is time normalized by the thermal life-time. The vertical axis is the laser detuning with respect to the cold cavity, normalized by the total cold-cavity linewidth.}
\end{figure}

Because the optical resonance shifts when the optical energy inside the cavity increases, it is necessary to solve equations~\ref{eq:5.15} and~\ref{eq:5.16} with a time dependent detuning ($ \Delta = \Delta (t) $),  as the example in figure~\ref{fig:DetTime}. In this function $ \Delta(t) $ varies from the blue side of the cold cavity ($ \Delta/\kappa>0 $), until a given stop point on the red side of the cold resonance ($ \Delta/\kappa<0 $). If this ramp is done too slowly, the solutions tend to show only static temperature and FC density. That is because the model doesn't take noise into account, hence there is nothing to perturb the solution out of its equilibrium.

On the other hand, if the ramp is done fast enough such that oscillations appear in the solutions, it typically causes artifacts on the solutions due to the variations of the stored optical energy being on the same time-scale of the phenomena of interest. To overcome this problem, the system of equations is solved multiple times, changing the stop detuning in each solution and storing only the data related to the times when $ \Delta $ is static (red line in fig.~\ref{fig:DetTime}). In this way the ramp of detuning can be done very quickly without affecting the final solution.

An important observation is that the $ \Delta $ values presented in figure~\ref{fig:DetTime} are relative to the cold cavity, i.e., without considering the non-linear resonance shift. As will be clear later in this section, the effective detuning is always positive with respect to the shifted resonance position. Also, we note that these numerical solutions presented here were obtained before proper characterization of the FC and thermal lifetimes of the samples, what invariably leads to deviations in quantitative results between model and experiment. Nevertheless, good agreement in the general behavior of the systems is achieved for both single and double cavity cases.

\subsection{Single cavity}
\label{sub:05.01.01}

\begin{figure}[ht!]
	\center
	\includegraphics[scale=1.0]{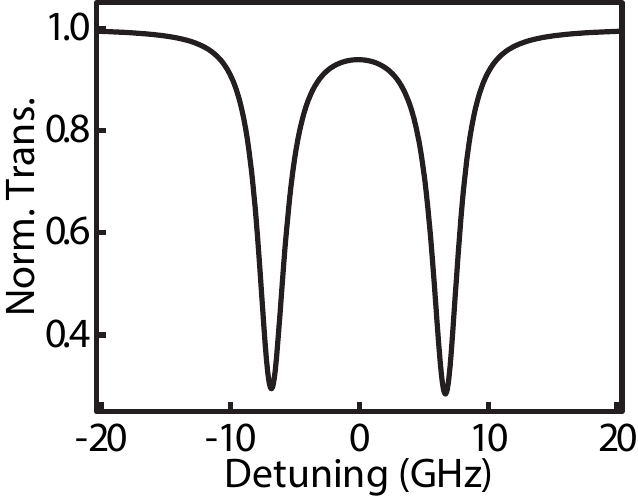}
	\caption[Numerical optical transmission spectrum of a single cavity with coupling of counter-propagating modes.]{\label{fig:NumSingCavSpec}\textbf{Numerical optical transmission spectrum of a single cavity with coupling of counter-propagating modes.} The coupling rate between modes is 13~GHz and the total optical decay rate is 2.5~GHz.}
\end{figure}

To reproduce the single cavity experiments the optical modes are modeled with a splitting due to coupling of counter propagating modes.
The expected optical transmission spectrum for a low power pump is shown in figure~\ref{fig:NumSingCavSpec}, with an optical decay rate $\kappa = 2\pi\times$2.5~GHz and a coupling rate between modes $\gcp =2\pi\times$13~GHz. The cavity-waveguide coupling rate ($ \kappa_e $) is 1.1~GHz and the central wavelength was set to approximately 1471~nm. The list of values for all parameters used in this case are listed in \cref{tab:params}.

\begin{figure}[ht!]
	\center
	\includegraphics[scale=1.0]{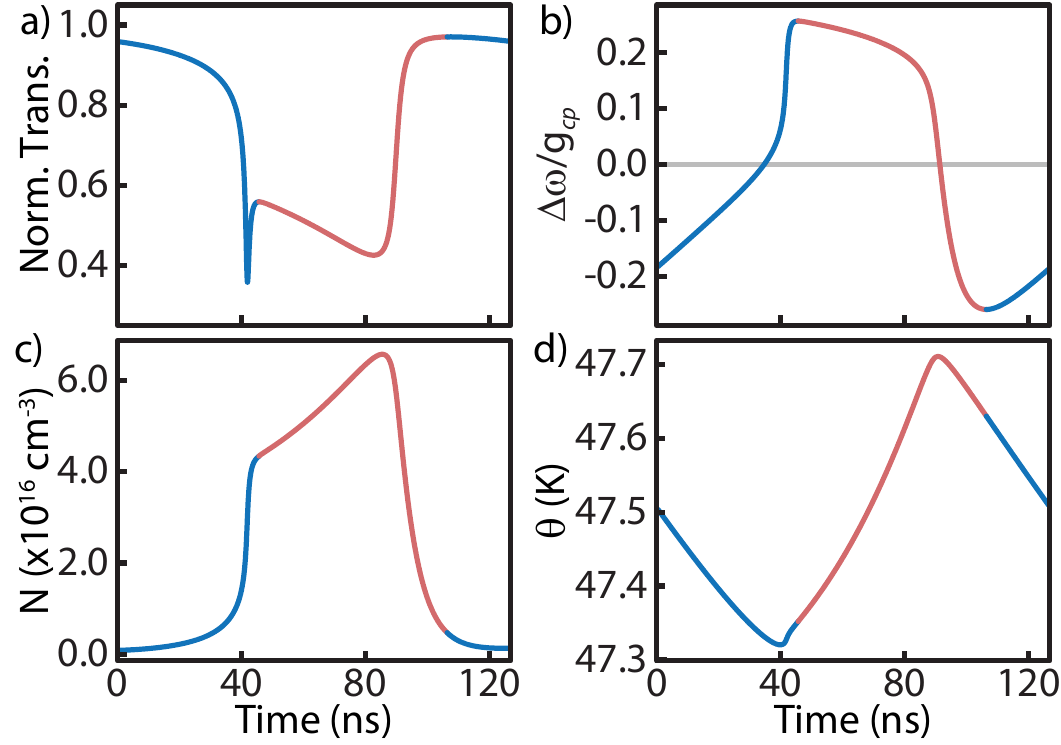}
	\caption[Numerical time-traces of a self-pulsing single optical cavity.]{\label{fig:FCTO-SimSingTT}\textbf{Numerical time-traces of a self-pulsing single optical cavity over one period.} a) Optical transmission normalized by the input power $ P_{in} $. b) Total resonance shift with respect to the mean resonance shift normalized to $ \gcp $. c) Free-carriers density. d) Temperature variation with respect to the cold-cavity equilibrium value. The colors indicate the regions where there is relative red and blue-shift of the resonance.}
\end{figure}

The equations are solved in~\mathematica~and, for a given final detuning value, result in time-traces like the ones shown in figure~\ref{fig:FCTO-SimSingTT}, which show the traces over one complete cycle of oscillation. By substituting the $ \theta(t) $ and $ N(t) $ solutions into equation~\ref{eq:5.14} and calculating the normalized time dependent transmission, the curve presented in figure~\ref{fig:FCTO-SimSingTT}a is obtained (analogous to Figure 1(b,c) in the main text). Additionally, Figure~\ref{fig:FCTO-SimSingTT}b shows the frequency shift of the resonances relative to the equilibrium position, which is given by
\begin{equation}
	\Delta\omega = \gtheta \theta(t) + \gn N(t) - \overline{\Delta\omega},\label{eq:5.17}
\end{equation}
\noindent
where $\overline{\Delta\omega} = \gtheta \overline{\theta}+\gn \overline{N}$ is the shift caused by the mean variation of both temperature and FC density, with the overbar indicating time average. Figures~\ref{fig:FCTO-SimSingTT}c and ~\ref{fig:FCTO-SimSingTT}d show the time behavior of FC and temperature variation over a single period. The colors indicate the regions where a relative blue-shift (blue) or red-shift (red) occurs, respectively.

\begin{figure}[ht!]
	\center
	\includegraphics[scale=1.0]{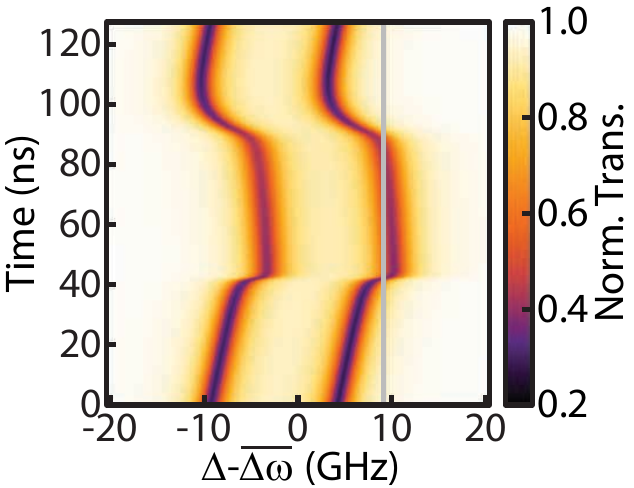}
	\caption[Numerical time-dependent optical transmission spectrogram.]{\label{fig:FCTO-SimSingTransSpecgram}\textbf{Numerical time-dependent optical transmission spectrogram.} The vertical line indicates the position of the pump laser frequency.}
\end{figure}

Observe that the resonance path reproduces exactly the trace presented in figure~\ref{fig:FCTO-SimSingTT}b, and now it is possible to see the two moments when the bluer resonance passes by the laser position, causing the minima in the pulse pattern. Also, observe that because the two resonances are of the same cavity, experiencing then the same shifts due to the FC and temperature variations, both resonances move exactly in phase, never changing their relative distance.

\begin{figure}[ht!]
	\center
	\includegraphics[scale=1.0]{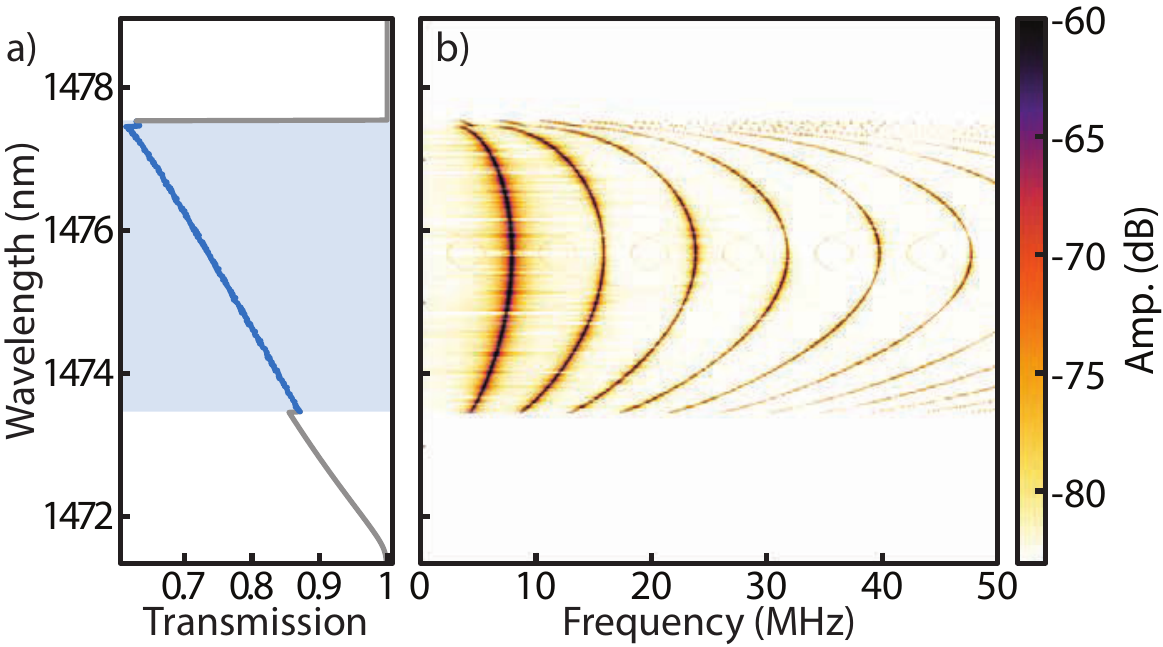}
	\caption[Numerical wavelenth dependent spectrogram of a self-pulsing single optical cavity.]{\label{fig:FCTO-SimSingSpecgram}\textbf{Numerical wavelength dependent spectrogram of a self-pulsing single optical cavity.} a) Normalized transmission of a detuning sweeping pump laser, with the typical triangular shape due to non-linear bi-stability~\cite{Barclay:05}. The blue region is where the self-pulsing occurs. b) Detuning dependent spectrogram of the transmitted self-pulsing light. The vertical scale follows the same scale in (a).}
\end{figure}

Finally, taking the Fourier transform of the optical transmission time-trace (fig. \ref{fig:FCTO-SimSingTT}a) for different static detuning positions produces a spectrogram, with the time axis substituted by a pump wavelength axis (fig.~\ref{fig:FCTO-SimSingSpecgram}). The pump wavelength dependent transmission (fig.~\ref{fig:FCTO-SimSingSpecgram}a) is obtained by taking the average of the oscillating transmission signal for each detuning. The pulses spectra present various harmonics, spanning up to the GHz scale, which is expected for the square-like pulses produced in this process. Note that the detuning for which oscillations start is very well marked by both an abrupt increase in the average transmitted signal and the appearance of the peaks in the spectrum.

\subsection{Coupled cavities}
\label{sub:05.01.02}

To obtain the equations for coupled cavities or for the case when a single cavity presents coupled modes, one must turn to the coupled mode theory of optical modes. This method uses the same initial equation, but now one for each mode coupled to each other, which in the rotating frame are given by
\begin{align}
	\frac{d a_{1}(t)}{d t}  &= -\mathfrak{i} \Delta_{1}\ a_{1}(t) - \frac{\kappa_{1}}{2}\ a_{1}(t) + \sqrt{\kappa_{e}}\ s_{in}s + \mathfrak{i}\frac{\gcc}{2}a_{2}(t)\hspace{0.5cm}\text{and}\label{eq:apA.6}\\
	\frac{d a_{2}(t)}{d t}  &= -\mathfrak{i} \Delta_{2}\ a_{2}(t) - \frac{\kappa_{2}}{2}\ a_{2}(t) + \mathfrak{i}\frac{\gcc}{2}a_{1}(t), \label{eq:apA.7}
\end{align}
\noindent
where $ \gcc $ is the coupling rate between modes 1 and 2, which can be either the modes of two cavities or two modes of the same cavity. Rigorously, this coupling rate can be estimated from the overlap of the modes and the energy and momentum conservation laws. Here we wrote the more general problem, with the two modes having different resonant frequencies, resulting in different detunings of the laser, and different intrinsic decay rates. Note that for a typical coupled cavity system $ \kappa_{2} $ doesn't include a contribution from $ \kappa_{e} $, as the second cavity only receives light through the first, and only the latter is coupled to the waveguide. This is not necessarily true for the case of coupling between two modes of the same cavity, e.g., when there is coupling of counter-propagating modes. Also, if there is coupling between different modes of the same cavity, because their coupling to the waveguide can be different, this must be considered and the equations modified. Nonetheless, we do not add such complications here because there is little contribution for the intended discussion.

\begin{figure}[ht!]
	\center
	\includegraphics[scale=1.0]{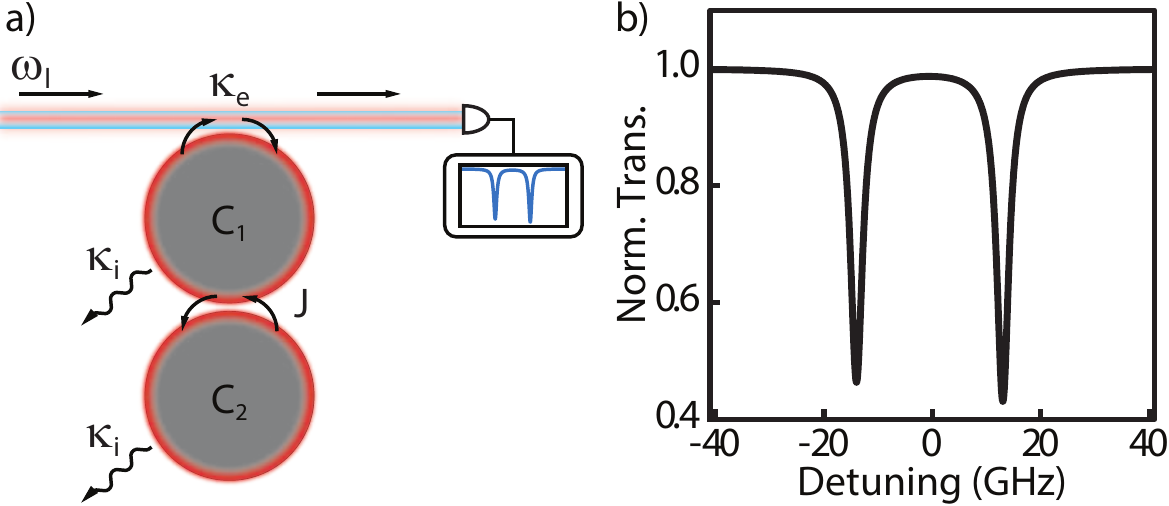}
	\caption[Schematics and transmission spectrum of a coupled cavity system model.]{\label{fig:NumDoubCavSpec}\textbf{Schematics and transmission spectrum of a coupled cavity system model.} a) Schematics of a coupled cavity system. The coupling between cavity modes is given by $ \gcc $. b) Optical transmission spectrum of a pair of coupled resonances. The coupling rate between modes is 28~GHz and the loaded optical decay rate is 2.5~GHz. The difference in extinction ratio is due to a mismatch in resonance frequency of 1~GHz between the two cavities.}
\end{figure}

As shown in figure~\ref{fig:NumDoubCavSpec}a, in this case one of the cavities is coupled directly to the waveguide 
(which will be called $ \cgg $), while the other ($ \cng $) only receives light that coupled into $ \cgg $. Figure~\ref{fig:NumDoubCavSpec}b shows the optical transmission spectrum for this system, as it would appear in a low input power measurement. The coupling rate between cavities is $ \gcc =2\pi\times $28~GHz, and the resonances are centered around 1471~nm with total linewidths $ \kappa = 2\pi\times$2.5~GHz.

Note that there is a slight asymmetry in the peaks extinction ratio; that is because a slight mismatch between the resonances of $ \cgg $ and $ \cng $ is taken into account. This mismatch was created by making the $ \cgg $ with resonance $ \omega_0 $ and $ \cng $ with resonance $ \omega_0-\delta_0 $, where $ \delta_0 $ is a positive constant, making the resonance of $ \cng $ redder than that of $ \cgg $. In this case the difference is $ \delta_0\approx2\pi\times$1~GHz. Besides this resonance frequency mismatch, the cavities are considered to be identical in terms of optical losses and material properties, as well as FC and thermal life-time.

Also, in the single cavity case it was demonstrated that the splitting due to coupling of counter-propagating modes doesn't add sensible information to the problem. At most it would cause split dips in the transmission time-trace, and only if the shift caused by temperature and FC is large enough for both peaks to pass by the pump laser frequency. Hence we didn't take this feature into account for the double cavity problem, even though the real device presents two kinds of splitting, a larger one for the coupling between cavities and two smaller ones for the coupling of counter-propagating modes in each cavity.

It is important to note that the two peaks in the spectrum of figure~\ref{fig:NumDoubCavSpec}b are not those of the individual resonances of $ \cgg $ and $ \cng $. Instead, these are what are commonly named super-modes of the coupled system, which are comprised of symmetric and anti-symmetric combinations of individual cavity modes, and their distance is related to the coupling rate $ \gcc $. Throughout this chapter we will refer to these resonances by coupled resonances, in this manner distinguishing them from the individual cavities' resonances.

In the double cavity case, the importance of reducing the number of equations for each cavity becomes apparent. The whole problem involves three equations per cavity, and the equation for the optical field amplitude ($ a(t) $) is complex, which in practice increases the number of equations to four per cavity, because any numerical solver will make $ a(t) = A(t)+\mathfrak{i} B(t) $ before starting to solve the system of equations. Removing the equation for $ a(t) $ by the adiabatic approximation (the field responds instantly to the temperature and FC variations) reduces the problem to a set of two real equations per cavity, making the problem much simpler and quicker to solve. The system of equations for two cavities is given by
\begin{align}
\frac{d N_{1}(t)}{d t} &= -\gamma_{FC} N_{1}(t) + \beta_{FC} U_{1}(t)^2\label{eq:5.18}\\
\frac{d \theta_{1}(t)}{d t} &= -\gamma_{th} \theta_{1}(t) + \beta_{th} \left( \kappa_{lin} + \sigma_{Si} v_g N_{1}(t) + \alpha_{TPA} U_{1}(t)^2 \right) U_{1}(t)^2\label{eq:5.19}\\
\frac{d N_{2}(t)}{d t} &= -\gamma_{FC} N_{2}(t) + \beta_{FC} U_{2}(t)^2\label{eq:5.20}\\
\frac{d \theta_{2}(t)}{d t} &= -\gamma_{th} \theta_{2}(t) + \beta_{th} \left( \kappa_{lin} + \sigma_{Si} v_g N_{2}(t) + \alpha_{TPA} U_{2}(t)^2 \right) U_{2}(t)^2,\label{eq:5.21}
\end{align}
\noindent
where $ N_{i} $, $ \theta_{i} $ and $ U_{i} $ are the FC density, temperature variation and intra-cavity energy of cavities $ \cgg $ and $ \cng $. Notice that there is no direct coupling between $ N $ and $ \theta $ of the two cavities, instead it is mediated exclusively by the optical field of the cavities, i.e. implicitly through $ U_{i} $, which are given by
\begin{align}
U_{1}(t) &=\frac{4 \left(4 \Delta_2^2+\kappa_2^2\right) \kappa_e P_{in}}{\left(4 \Delta_1^2+\kappa_1^2\right) \left(4 \Delta_2^2+\kappa_2^2\right)+\gcc^2 \left(2 \kappa_1 \kappa_2-8 \Delta_1 \Delta_2\right)+\gcc^4}\label{eq:5.22}\\
U_{2}(t) &= \frac{4 \gcc^2 \kappa_e P_{in}}{\left(4 \Delta_1^2+\kappa_1^2\right) \left(4 \Delta_2^2+\kappa_2^2\right)+\gcc^2 \left(2 \kappa_1 \kappa_2-8 \Delta_1 \Delta_2\right)+\gcc^4},\label{eq:5.23}
\end{align}
\noindent
where $ \Delta_{1} = \Delta_{0} + \gtheta \theta_{1} + \gn N_{1} $ and $ \Delta_{2} = \Delta_{0} + \gtheta \theta_{2} + \gn N_{2}+\delta_0 $~are the effective detunings due to the variation of $ \theta $ and $ N $ in each cavity, with $ \Delta_{0} $ being the detuning with respect to the unperturbed resonance of $ \text{C}_{1} $; and $ \kappa_{1,2} = \kappa+\alpha_{2}N_{1,2}$  are the decay rates of each cavity, contemplating both linear and non-linear contributions.


\end{document}